\documentclass[10pt,conference]{IEEEtran}
\IEEEoverridecommandlockouts

\usepackage{cite}
\usepackage{amsmath,amssymb,amsfonts}
\usepackage{graphicx}
\usepackage{textcomp}
\usepackage{xcolor}
\usepackage{colortbl}
\usepackage{booktabs}
\usepackage{array}
\usepackage{url}
\usepackage{tikz}
\usetikzlibrary{shapes.geometric,arrows.meta,positioning,calc,backgrounds,fit,shadows,decorations.pathreplacing}
\usepackage{pgfplots}
\pgfplotsset{compat=1.18}
\usepackage[hidelinks]{hyperref}
\hypersetup{hypertexnames=false}
\usepackage{qrcode}
\usepackage{changepage}
\usepackage{float}
\usepackage[most]{tcolorbox}  

\definecolor{qblue}{RGB}{30,80,160}
\definecolor{qpurple}{RGB}{120,40,140}
\definecolor{qgreen}{RGB}{40,140,90}
\definecolor{qorange}{RGB}{220,120,30}
\definecolor{qgrey}{RGB}{90,90,90}
\definecolor{qlight}{RGB}{235,240,250}
\definecolor{mF1}{HTML}{2E7D8C}\definecolor{mAP}{HTML}{E0A33E}\definecolor{mAUC}{HTML}{AEB6BE}
\definecolor{corrIceFill}{HTML}{E8F2F9}\definecolor{corrIceLine}{HTML}{3E7CA6}\definecolor{corrIceText}{HTML}{27506B}
\definecolor{wfq}{HTML}{3E7CA6}\definecolor{wfc}{HTML}{C77A30}
\definecolor{corrItFill}{HTML}{E3F1ED}\definecolor{corrItLine}{HTML}{2F7D8C}\definecolor{corrItText}{HTML}{1F5560}
\definecolor{corrCncFill}{HTML}{F3F0E9}\definecolor{corrCncText}{HTML}{6B6356}

\def\BibTeX{{\rm B\kern-.05em{\sc i\kern-.025em b}\kern-.08em
    T\kern-.1667em\kern-.125emE\kern-.125emX}}

\begin{document}


\onecolumn
\thispagestyle{empty}
\vspace*{1.2cm}

\begin{center}
{\large\bfseries From IceCube to \textit{IT-Sphere}:\\[2pt]
A Hybrid Quantum--Classical GNN for Banking IT Root Cause Analysis\par}
\vspace{6pt}
{\normalsize Antonio Greco, Riccardo Paoletti, Roberto Cappuccio, Mario Onorato\par}
\end{center}

\vspace{1.0cm}

\begin{tcolorbox}[enhanced, colback=qlight, colframe=qblue, boxrule=0.8pt,
  arc=2pt, left=8pt, right=8pt, top=6pt, bottom=6pt,
  title={\bfseries IEEE Copyright Notice}, fonttitle=\normalsize,
  coltitle=white, colbacktitle=qblue]
\copyright~2026 IEEE. Personal use of this material is permitted. Permission
from IEEE must be obtained for all other uses, in any current or future media,
including reprinting/republishing this material for advertising or promotional
purposes, creating new collective works, for resale or redistribution to
servers or lists, or reuse of any copyrighted component of this work in other
works.
\end{tcolorbox}

\vspace{0.5cm}

\begin{tcolorbox}[enhanced, colback=qgreen!6, colframe=qgreen, boxrule=0.8pt,
  arc=2pt, left=8pt, right=8pt, top=6pt, bottom=6pt,
  title={\bfseries Publication status}, fonttitle=\normalsize,
  coltitle=white, colbacktitle=qgreen]
Accepted for publication in the \emph{Proceedings of the 2026 IEEE
International Conference on Quantum Computing and Engineering (QCE26)},
Metro Toronto Convention Centre, Toronto, ON, Canada, 13--18 September 2026.
Paper ID POS2-1600, Poster Track. DOI: to appear.

\vspace{4pt}
\footnotesize This is the author's accepted version. It is not the
IEEE-published version of record.
\end{tcolorbox}

\vspace{0.5cm}

\begin{tcolorbox}[enhanced, colback=qgrey!5, colframe=qgrey, boxrule=0.8pt,
  arc=2pt, left=8pt, right=8pt, top=6pt, bottom=6pt,
  title={\bfseries Cite as}, fonttitle=\normalsize,
  coltitle=white, colbacktitle=qgrey]
A.~Greco, R.~Paoletti, R.~Cappuccio, and M.~Onorato, ``From IceCube to
IT-Sphere: A Hybrid Quantum--Classical GNN for Banking IT Root Cause
Analysis,'' in \emph{2026 IEEE International Conference on Quantum Computing
and Engineering (QCE)}, Toronto, ON, Canada, 2026, to be published.

\vspace{8pt}
\hrule
\vspace{8pt}
\begin{verbatim}
@inproceedings{greco2026hqrca,
  author    = {Greco, Antonio and Paoletti, Riccardo and
               Cappuccio, Roberto and Onorato, Mario},
  title     = {From {IceCube} to {IT-Sphere}: A Hybrid
               Quantum--Classical {GNN} for Banking {IT}
               Root Cause Analysis},
  booktitle = {2026 IEEE International Conference on Quantum
               Computing and Engineering (QCE)},
  address   = {Toronto, ON, Canada},
  year      = {2026},
  note      = {to be published}
}
\end{verbatim}
\end{tcolorbox}

\clearpage

\makeatletter
\twocolumn[
\begin{@twocolumnfalse}
\vspace*{-18pt}

\begin{center}
{\fontsize{24pt}{28pt}\selectfont%
From IceCube to \textit{IT-Sphere}:\\[2pt]
A Hybrid Quantum--Classical GNN for Banking IT Root Cause Analysis\par}
\end{center}

\vspace{2pt}

\begin{center}
{\normalsize Antonio Greco$^{*\dagger}$, Riccardo Paoletti$^{*\dagger}$, Roberto Cappuccio$^{*\dagger}$, Mario Onorato$^{\ddagger}$}
\end{center}

\vspace{1pt}

\begin{center}
{\small $^{*}$\textit{Department of Physical Sciences, Earth and Environment (DSFTA), University of Siena}, Siena, Italy~$\bullet$~$^{\dagger}$\textit{Istituto Nazionale di Fisica Nucleare (INFN)}, Pisa, Italy~$\bullet$~$^{\ddagger}$\textit{IBM Consulting---Promontory, IBM Industry Diamond, Banking \& Financial Markets}, Milan, Italy

a.greco19@student.unisi.it~$\bullet$~riccardo.paoletti@unisi.it~$\bullet$~roberto.cappuccio@pi.infn.it~$\bullet$~monorato@promontory.com}
\end{center}

\vspace{2pt}

\noindent\hspace{2em}\begin{minipage}{\dimexpr\textwidth-4em\relax}
\small\textbf{Abstract---We present Hybrid Quantum Root Cause Analysis (HQ-RCA), an industrially grounded workflow for root cause analysis in banking IT operations, built on a hybrid Quantum Graph Neural Network (QGNN): the classical backbone of DynEdge (the IceCube neutrino-reconstruction GNN, which we call \emph{standalone DynEdge}), with its classification head replaced by a Variational Quantum Circuit (VQC). On 13~months of anonymised IT data (13k alarm clusters) from a major European bank, its hybrid QGNN matches standalone DynEdge --- the strongest classical baseline --- on $F_1$, while standalone DynEdge leads the ranking metrics. A readout-sensitivity and layout-robustness study, analysed via Dimensional Expressivity Analysis (DEA), shows that the effective parameter dimensionality (rank) of the quantum observable has no measurable correlation with $F_1$; we therefore keep the simplest readout $\langle Z_0\rangle$ (the Pauli-$Z$ expectation on the first qubit), which in the deployed layout is rank-1, collapsing optimisation to a 1-D problem solvable by a gradient-free grid scan. Execution on IBM Heron r2 (no error mitigation) shows this gradient-free readout is executable on NISQ hardware after threshold recalibration.}
\end{minipage}

\vspace{4pt}
\end{@twocolumnfalse}]
\makeatother
\vspace*{-15pt}
\section{Introduction}
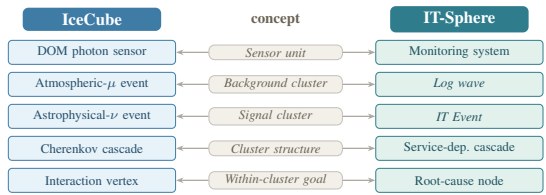
\begin{figure}[!b]
\centering\resizebox{0.80\columnwidth}{!}{%
\begin{tikzpicture}[
  font=\sffamily,
  ice/.style={rounded corners=2pt, fill=corrIceFill, draw=corrIceLine, line width=0.5pt,
              text=corrIceText, align=center, inner sep=2pt, minimum height=4.5mm, text width=30mm, font=\scriptsize},
  it/.style={rounded corners=2pt, fill=corrItFill, draw=corrItLine, line width=0.5pt,
              text=corrItText, align=center, inner sep=2pt, minimum height=4.5mm, text width=30mm, font=\scriptsize},
  cnc/.style={rounded corners=3pt, fill=corrCncFill, draw=corrCncText!40, line width=0.4pt,
              text=corrCncText, align=center, inner sep=1.5pt, text width=24mm, font=\scriptsize\itshape},
  link/.style={-{Stealth[length=4pt]}, draw=corrCncText!55, line width=0.7pt},
]
\node[rounded corners=3pt, fill=corrIceLine, text=white, font=\small\bfseries, inner sep=3pt, minimum width=26mm] at (0,0.62) {IceCube};
\node[rounded corners=3pt, fill=corrItLine, text=white, font=\small\bfseries, inner sep=3pt, minimum width=26mm] at (6.9,0.62) {IT-Sphere};
\node[text=corrCncText, font=\footnotesize\bfseries] at (3.45,0.62) {concept};
\foreach \i/\L/\R/\C in {0/{DOM photon sensor}/{Monitoring system}/{Sensor unit},
                         1/{Atmospheric-$\mu$ event}/{\textit{Log wave}}/{Background cluster},
                         2/{Astrophysical-$\nu$ event}/{\textit{IT Event}}/{Signal cluster},
                         3/{Cherenkov cascade}/{Service-dep.\ cascade}/{Cluster structure},
                         4/{Interaction vertex}/{Root-cause node}/{Within-cluster goal}}{
  \pgfmathsetmacro\yy{-\i*0.60}
  \node[ice] (l\i) at (0,\yy) {\L};
  \node[it]  (r\i) at (6.9,\yy) {\R};
  \node[cnc] (c\i) at (3.45,\yy) {\C};
  \draw[link] (c\i.west) -- (l\i.east);
  \draw[link] (c\i.east) -- (r\i.west);
}
\end{tikzpicture}}
\vspace{-6pt}
\caption{IceCube\,$\leftrightarrow$\,IT-Sphere structural correspondence: each centre concept links a neutrino-detector notion (left) to its banking-IT counterpart (right) --- a structural analogy, not a task identity.}
\label{fig:corr}
\end{figure}
Modern banks run continuous IT monitoring that emits a steady stream of alarms; industrial event-correlation tools cluster them by recurring patterns but are not primarily designed to reconstruct the causal chain across interconnected services --- a diagnosis still made by hand. We cast root cause analysis (RCA) as a graph-reconstruction problem inspired by IceCube neutrino-event reconstruction. Each alarm cluster is a graph: a \emph{log wave} (background cluster, analogous to an atmospheric muon) or a rare \emph{IT Event} (an incident hiding, by assumption, a single \emph{root-cause} alarm, analogous to a neutrino vertex). We term the banking IT ecosystem the \emph{IT-Sphere}, the analogue of IceCube's detector volume (Fig.~\ref{fig:corr}) --- a structural analogy, not system identity. Clusters are weakly labelled by a heuristic root-cause score, the labelling rule validated by domain experts. On a 2{,}000-graph subset (400-graph golden set; one train/test split and CV at a fixed seed (42), identical for every model) we benchmark Random Forest (RF), XGBoost (XGB), the IceCube DynEdge run end-to-end (\emph{standalone DynEdge}), and the hybrid QGNN, which reuses that backbone with a VQC in place of the head (\S\ref{sec:method}; Fig.~\ref{fig:f1}).

\section{Method and architectural selection}\label{sec:method}
\begin{figure}[!t]
\centering
\resizebox{0.98\columnwidth}{!}{%
\begin{tikzpicture}[
  font=\sffamily,
  qn/.style={rounded corners=2pt, draw=wfq, fill=wfq!8, text=wfq!55!black, line width=0.6pt, inner sep=3pt, minimum height=7mm, align=center, font=\footnotesize},
  cn/.style={rounded corners=2pt, draw=wfc, fill=wfc!10, text=wfc!50!black, line width=0.6pt, inner sep=3pt, minimum height=7mm, align=center, font=\footnotesize},
  ar/.style={-{Stealth[length=5pt]}, line width=0.8pt, draw=black!55},
  tl/.style={font=\scriptsize\itshape, text=black!65},
]
\node[qn] (th) {$\boldsymbol{\theta}$};
\node[qn, right=12mm of th] (psi) {$|\psi(\boldsymbol{\theta},x_v)\rangle$};
\node[qn, right=12mm of psi] (z0) {$\langle Z_0\rangle$};
\node[cn, right=13mm of z0] (py) {$\mathbb{P}(y|x_v)$};
\node[cn, right=12mm of py] (yh) {$\hat{y}_v$};
\node[cn, right=10mm of yh] (f1) {$F_1$};
\draw[ar] (th)  -- node[tl,above]{circuit}            (psi);
\draw[ar] (psi) -- node[tl,above]{meas.}              (z0);
\draw[ar] (z0)  -- node[tl,above]{softmax}            (py);
\draw[ar] (py)  -- node[tl,above]{$\tau$-thresh}      (yh);
\draw[ar] (yh)  -- node[tl,above]{P/R}                (f1);
\draw[decorate,decoration={brace,amplitude=4pt,mirror},draw=wfq,line width=0.6pt]
  ([yshift=-5pt]th.south west) -- node[below=6pt,font=\footnotesize\bfseries,text=wfq]{quantum} ([yshift=-5pt]z0.south east);
\draw[decorate,decoration={brace,amplitude=4pt,mirror},draw=wfc,line width=0.6pt]
  ([yshift=-5pt]py.south west) -- node[below=6pt,font=\footnotesize\bfseries,text=wfc]{classical} ([yshift=-5pt]f1.south east);
\end{tikzpicture}}
\vspace{-6pt}
\caption{HQ-RCA end-to-end workflow: a quantum block ($\boldsymbol{\theta}{\to}\langle Z_0\rangle$) feeding a classical block ($\langle Z_0\rangle{\to}F_1$). AP and ROC-AUC are read off $\mathbb{P}(y|x_v)$ directly (threshold-free).}
\label{fig:pipeline}
\end{figure}
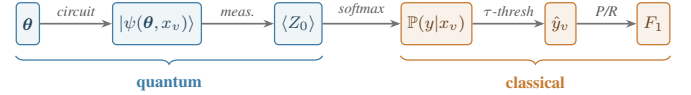
We reuse DynEdge~\cite{abbasi2022}, the GNN built for IceCube neutrino reconstruction, as the classical backbone --- it handles variable-topology graphs from Union--Find grouping. The VQC~\cite{cerezo2021} is built from an \emph{encoding} $E$ and a parametric \emph{ansatz} $A$.
The end-to-end pipeline is: (i)~Union--Find grouping and principal component analysis (PCA) on the node features ($8{\to}7$, 95\% variance retained);
(ii)~DynEdge --- four EdgeConv blocks (graph edge-convolution layers that update each node from its neighbours) producing a node embedding $h_v\!\in\!\mathbb{R}^{1031}$, with $1031{=}7{+}4{\times}256$;
(iii)~linear layer$+\tanh$ activation: $\mathbb{R}^{1031}\!\to\!x_v\!\in\!(-1,+1)^{4}$;
(iv)~4-qubit VQC encodes $x_v$ via $E$, applies the ansatz $A$, and measures the single-qubit baseline observable $\langle Z_0\rangle$ (analysed in \S\ref{sec:dea});
(v)~linear head outputs the root-cause probability $\mathbb{P}(y|x_v,\boldsymbol{\theta})=\mathrm{softmax}(\mathbf{w}\,\langle Z_0\rangle(x_v,\boldsymbol{\theta})+\mathbf{b})_y$, where $\langle Z_0\rangle\!\in\![-1,+1]$ is the expectation of the
Pauli observable $Z_0\!=\!Z\!\otimes\!I^{\otimes 3}$ on the first qubit, $\mathbf{w},\mathbf{b}\!\in\!\mathbb{R}^{2}$
are trainable head parameters, and
$y\!\in\!\{0,1\}$ is the class label. $F_1$ also depends on the data, preprocessing and the trained classical components; the map
$F_1\!:\!\mathbb{R}^{12}\!\times\![0,1]\!\to\![0,1]$, $(\boldsymbol{\theta},\tau)\!\mapsto\!F_1$, is the pipeline of Fig.~\ref{fig:pipeline} ($\tau$: decision threshold, fixed at $\tau{=}0.5$ in simulation, recalibrated on hardware, \S\ref{sec:hw}).

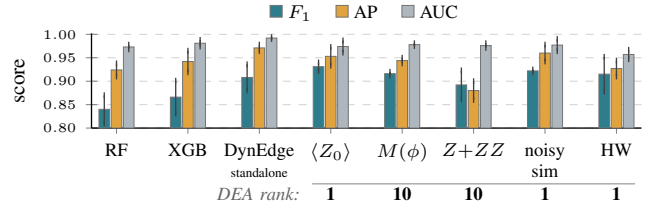
\begin{figure}[t]
\centering
\begin{tikzpicture}
\begin{axis}[
  width=\columnwidth, height=0.32\columnwidth,
  ybar=0.6pt, bar width=4pt, xmin={[normalized]-0.35}, xmax={[normalized]7.35},
  ymin=0.80, ymax=1.002, ytick={0.80,0.85,0.90,0.95,1.00},
  yticklabel style={font=\tiny,/pgf/number format/fixed,/pgf/number format/fixed zerofill,/pgf/number format/precision=2},
  ymajorgrids, grid style={gray!35,dashed,line width=0.35pt},
  ylabel={score}, ylabel style={font=\footnotesize},
  symbolic x coords={RF,XGB,DynEdge,Z0,Mphi,fullP,noisy,HW}, xtick=data,
  xticklabels={RF,XGB,{DynEdge\\{\tiny standalone}},$\langle Z_0\rangle$,$M(\phi)$,$Z{+}ZZ$,{noisy\\sim},HW},
  xticklabel style={font=\scriptsize,align=center}, tick align=outside,
  axis lines=left, axis line style={-}, line width=0.5pt, clip=false,
  legend style={at={(0.5,1.0)},anchor=south,legend columns=3,draw=none,font=\scriptsize,
                /tikz/every even column/.append style={column sep=7pt}},
  legend image code/.code={\draw[#1,draw=black!55,line width=0.3pt] (0cm,-0.045cm) rectangle (0.16cm,0.11cm);},
]
\addplot[fill=mF1,draw=black!55,line width=0.3pt,
  error bars/.cd,y dir=both,y explicit,
  error bar style={black!75,line width=0.5pt},error mark options={black!75,line width=0.5pt,mark size=1.1pt}]
  coordinates {(RF,0.840)+-(0,0.030) (XGB,0.866)+-(0,0.035) (DynEdge,0.908)+-(0,0.028)
    (Z0,0.931)+-(0,0.009) (Mphi,0.916)+-(0,0.004) (fullP,0.892)+-(0,0.031)
    (noisy,0.922)+-(0,0.003) (HW,0.915)+-(0,0.037)};
\addplot[fill=mAP,draw=black!55,line width=0.3pt,
  error bars/.cd,y dir=both,y explicit,
  error bar style={black!75,line width=0.5pt},error mark options={black!75,line width=0.5pt,mark size=1.1pt}]
  coordinates {(RF,0.924)+-(0,0.014) (XGB,0.942)+-(0,0.023) (DynEdge,0.971)+-(0,0.007)
    (Z0,0.953)+-(0,0.020) (Mphi,0.944)+-(0,0.006) (fullP,0.880)+-(0,0.020)
    (noisy,0.960)+-(0,0.018) (HW,0.927)+-(0,0.017)};
\addplot[fill=mAUC,draw=black!55,line width=0.3pt,
  error bars/.cd,y dir=both,y explicit,
  error bar style={black!75,line width=0.5pt},error mark options={black!75,line width=0.5pt,mark size=1.1pt}]
  coordinates {(RF,0.973)+-(0,0.005) (XGB,0.981)+-(0,0.007) (DynEdge,0.992)+-(0,0.002)
    (Z0,0.974)+-(0,0.013) (Mphi,0.978)+-(0,0.003) (fullP,0.976)+-(0,0.005)
    (noisy,0.977)+-(0,0.013) (HW,0.957)+-(0,0.010)};
\legend{$F_1$,AP,AUC}
\node[anchor=base east,font=\scriptsize\itshape,text=black!60] at ([yshift=-27pt,xshift=-8pt]axis cs:Z0,0.80) {DEA rank:};
\node[anchor=base,font=\scriptsize\bfseries] at ([yshift=-27pt]axis cs:Z0,0.80) {1};
\node[anchor=base,font=\scriptsize\bfseries] at ([yshift=-27pt]axis cs:Mphi,0.80) {10};
\node[anchor=base,font=\scriptsize\bfseries] at ([yshift=-27pt]axis cs:fullP,0.80) {10};
\node[anchor=base,font=\scriptsize\bfseries] at ([yshift=-27pt]axis cs:noisy,0.80) {1};
\node[anchor=base,font=\scriptsize\bfseries] at ([yshift=-27pt]axis cs:HW,0.80) {1};
\draw[black!45,line width=0.4pt]
  ([yshift=-21pt,xshift=-7pt]axis cs:Z0,0.80) -- ([yshift=-21pt,xshift=7pt]axis cs:HW,0.80);
\end{axis}
\end{tikzpicture}
\vspace{-9pt}
\caption{$F_1$, AP, ROC-AUC for all models (2{,}000-graph subset; AP chance ${\approx}0.17$); DEA rank below the quantum readouts. Error bars: $\sigma$, three held-out subsets (HW: five folds).}
\label{fig:f1}
\end{figure}

\emph{Architectural selection (encoding--ansatz sweep).} Two baselines motivate the search: (0a) RY (Y-rotation) encoding, one $R_y$ per qubit, + StronglyEntanglingLayer${\times}1$ (12 PQC params); (0b) a one-layer hardware-efficient embedding (HEE1) with an SO(4) two-qubit gate~\cite{elhag2024} (24 params). (0a) outperforms (0b), though the two differ in both axes. To disentangle the two axes we run an \emph{L-shaped sweep} of 13 distinct configurations: (1) encoding fixed (RY, 9 ansatz variants) $\to$ best ansatz RandomLayers; (2) ansatz fixed (RandomLayers, 5 encodings incl.\ RY) $\to$ best encoding HEE2 (two HEE layers). Both axes strongly affect performance; \emph{HEE2+RandomLayers} (12 VQC params) is selected (best $F_1{=}0.931$, AP${=}0.953$, AUC${=}0.974$; Fig.~\ref{fig:f1}). \emph{Data scaling} (2k$\to$13k): $F_1{=}0.915{\pm}0.017$, $\mathrm{AP}{=}0.947{\pm}0.020$ and ROC-AUC${=}0.973{\pm}0.013$, each within $1\sigma$ of its 2k value (Fig.~\ref{fig:f1}) --- consistent at scale.

\section{Effective dimensionality and readout choice}\label{sec:dea}
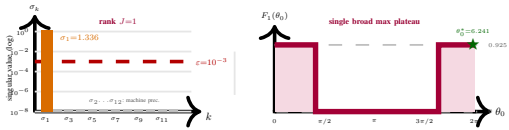
\begin{figure}[b]
\centering
\resizebox{0.76\columnwidth}{!}{%
\begin{tikzpicture}[scale=0.26,transform shape]

\draw[->,thick] (0,0) -- (5.0,0) node[right,font=\small\bfseries,yshift=-4pt]{$k$};
\draw[->,thick] (0,0) -- (0,2.9) node[above,font=\scriptsize]{$\sigma_k$};
\node[font=\scriptsize,rotate=90] at (-0.75,1.3) {singular value (log)};
\foreach \exp/\ypos in {0/2.4,-2/1.8,-4/1.2,-6/0.6,-8/0}{
  \draw[thin] (-0.05,\ypos) -- (0.05,\ypos);
  \node[font=\tiny,anchor=east] at (-0.05,\ypos) {$10^{\exp}$};
}
\foreach \ypos in {0,0.6,1.2,1.8,2.4}{
  \draw[gray!20,thin] (0,\ypos) -- (4.7,\ypos);
}
\draw[dashed,red!70!black,thick] (0,1.5) -- (4.7,1.5)
   node[right,font=\scriptsize,red!70!black]{$\varepsilon{=}10^{-3}$};
\fill[orange!85!black] (0.2,0) rectangle (0.55,2.44);
\node[font=\scriptsize\bfseries,orange!85!black,anchor=west] at (0.64,2.2) {$\sigma_1{=}1.336$};
\foreach \i in {2,...,12}{
  \pgfmathsetmacro{\xpos}{0.2+(\i-1)*0.35}
  \pgfmathsetmacro{\xposEnd}{\xpos+0.25}
  \fill[gray!50] (\xpos,0) rectangle (\xposEnd,0.06);
}
\foreach \i in {1,3,5,7,9,11}{
  \pgfmathsetmacro{\xpos}{0.2+(\i-1)*0.35+0.125}
  \node[font=\tiny] at (\xpos,-0.25) {$\sigma_{\i}$};
}
\node[font=\scriptsize\bfseries,purple!80!black,align=center] at (2.5,2.7) {rank $J{=}1$};
\node[font=\tiny,gray!70!black,anchor=west] at (1.5,0.32) {$\sigma_2{\dots}\sigma_{12}$: machine prec.};

\draw[gray!20,thin] (6.0,0.5) -- (6.0,1.8);

\begin{scope}[shift={(7.2,0)}]
  \draw[->,thick] (0,0) -- (6.5,0) node[right,font=\small\bfseries]{$\theta_0$};
  \draw[->,thick] (0,0) -- (0,2.6) node[above,font=\scriptsize]{$F_1(\theta_0)$};
  \draw[dashed,gray!50,thin] (0,2.0) -- (6.3,2.0);
  \node[font=\tiny,gray!70!black,anchor=west] at (6.3,2.0) {$0.925$};
  \draw[dashed,gray!50,thin] (0,0) -- (6.3,0);
  \node[font=\tiny,gray!70!black,anchor=west] at (6.3,0) {$0$};
  \foreach \xt/\lab in {0/{0},1.5/{\pi/2},3.0/{\pi},4.5/{3\pi/2},6.0/{2\pi}} {
    \draw[thick] (\xt,-0.05) -- (\xt,0.05);
    \node[font=\tiny,anchor=north] at (\xt,-0.10) {$\lab$};
  }
  \fill[purple!15] (0,0) rectangle (1.230,2.0);
  \fill[purple!15] (4.923,0) rectangle (6.0,2.0);
  \draw[purple!85!black,line width=1.2pt]
    (0,2.0) -- (1.230,2.0) -- (1.230,0) -- (4.923,0) -- (4.923,2.0) -- (6.0,2.0);
  \node[font=\large,green!40!black] at (5.96,2.0) {$\bigstar$};
  \node[font=\tiny\bfseries,green!40!black,anchor=south] at (5.96,2.15) {$\theta_0^{*}{=}6.241$};
  \node[font=\scriptsize\bfseries,purple!80!black,anchor=south,align=center] at (3.0,2.45) 
       {single broad max plateau};
\end{scope}
\end{tikzpicture}}
\caption{Left: DEA SVD spectrum (log) --- only $\sigma_1{=}1.336$ exceeds $\varepsilon{=}10^{-3}$. Right: $F_1(\theta_0)$ from a 40-point grid scan, a single broad plateau ($\sim$37\% of period) peaking at $\theta_0^{*}{=}6.241$~rad (\textcolor{green!40!black}{$\bigstar$}).}
\label{fig:svdspec}
\end{figure}

\emph{Effective dimensionality of the $\langle Z_0\rangle$ readout.} With the architecture fixed (HEE2{+}RandomLayers) and the baseline readout $\langle Z_0\rangle$, we ask how many of the 12 parameters the readout actually uses. Dimensional Expressivity Analysis (DEA) answers this through the thin SVD of the empirical \emph{observable} Jacobian (not the state Jacobian of~\cite{funcke2021}) $J{=}\partial\langle Z_0\rangle/\partial\boldsymbol{\theta}$ from 200 uniformly sampled inputs (a $200{\times}12$ matrix); the rank counts singular values above a cutoff $\varepsilon{=}10^{-3}$ (well above the noise floor). Here $J$ is rank-1: $\sigma_1{=}1.336$ while $\sigma_2,\dots,\sigma_{12}$ sit at machine precision, ${\sim}10^{-16}$ (Fig.~\ref{fig:svdspec}, left). Since $\mathbb{P}(y|x,\boldsymbol{\theta})$ depends on $\boldsymbol{\theta}$ only through the scalar $\langle Z_0\rangle$, the Fisher rank equals $\mathrm{rank}(J)$ exactly. The whole optimisation thus collapses onto one effective direction: $F_1$ as a function of the lone effective angle $\theta_0$ --- traced by a 40-point grid scan over one period --- is a $2\pi$-periodic step function with a single broad plateau (Fig.~\ref{fig:svdspec}, right). Fine-tuning $\theta_0$ alone --- the other eleven frozen at their trained values --- reaches $F_1{=}0.925$ at $\theta_0^{*}{=}6.241$~rad; with $\theta_0$ re-initialised at random, retraining still reaches $F_1{=}0.911$ (start-dependent).

\emph{Effect of a richer readout.} This rank-1 collapse is specific to $\langle Z_0\rangle$, so we test richer readouts. On the same circuit, splits and hyperparameters we compare three readouts: the baseline $\langle Z_0\rangle$ (one Pauli expectation, no extra parameters); a parametric single-qubit readout $M(\phi){=}\sum_i\phi_i\langle Z_i\rangle$ (4 parameters); and a two-local Z readout $M(\phi,\xi){=}\sum_i\phi_i\langle Z_i\rangle+\sum_{i<j}\xi_{ij}\langle Z_iZ_j\rangle$ (single and two-qubit Z terms, 10 parameters). The richer readouts raise the effective rank but not $F_1$: $\langle Z_0\rangle$ gives the best held-out $F_1$ ($0.931{\pm}0.009$, rank 1) and the smallest variance, against $0.916{\pm}0.004$ for $M(\phi)$ and $0.892{\pm}0.031$ for the two-local Z readout (both rank 10); AP and ROC-AUC stay comparable across the three (Fig.~\ref{fig:f1}). To isolate what the quantum part adds, we replace the VQC with a classical head of the same parameter count (Linear$+\tanh\to$scalar, not DynEdge's native head). At $\tau{=}0.5$ it reaches $F_1$ $0.876{\pm}0.063$ (AP/ROC-AUC $0.961/0.986$); recalibrating only its threshold ($\tau^{*}{\approx}0.76$, 5-fold) lifts $F_1$ to $0.907{\pm}0.020$ --- still below the VQC's $0.931{\pm}0.009$ --- so its $F_1$ deficit is not merely calibration. The classical head instead leads the threshold-free ranking metrics (AP/ROC-AUC). The quantum head is feasible, not decisively advantageous. \emph{Layout dependence of the rank.} We test whether rank-1 is intrinsic to the architecture or specific to this layout. Sweeping the RandomLayers seed over its first 21 values (0--20), the $\langle Z_0\rangle$ rank ranges from 0 to 9 (mean 3.3) --- rank-1 is the exception, not the rule. Yet across these 21 layouts, $F_1$ stays high and no correlation between $F_1$ and the rank is observed (Pearson $-0.08$, $p{=}0.73$). We therefore adopt $\langle Z_0\rangle$: rank-1 at the deployed layout, hence 1-D-optimisable and gradient-free on hardware.

\section{NISQ hardware execution}\label{sec:hw}
Under simulated depolarising noise, $F_1$ stays ${\geq}0.90$ up to a 5\% two-qubit gate-error rate ($p_{2q}{=}0.05$); a shot-noise sweep at $p_{2q}{=}0.01$ fixes 200 shots/node: $F_1{=}0.918$, near the $0.922$ plateau at ${\geq}500$ shots. With $\theta_0$ at its \S\ref{sec:dea} optimum, it runs on \texttt{ibm\_fez} (IBM Heron r2, no error mitigation) in ${\sim}2$~min. Device noise shifts the score distribution: at $\tau{=}0.5$ the model over-predicts and $F_1$ falls to $0.587$, whereas the threshold-invariant ranking metrics are largely preserved (AP $0.927$, ROC-AUC $0.957$, vs $0.960/0.977$ in simulation). Recalibrating only the threshold ($\tau^{*}{\approx}0.68$), validated by 5-fold cross-validation, recovers $F_1{=}0.915{\pm}0.037$ (matching the noisy-simulator ${\sim}0.92$) --- noise shifts the optimal threshold more than class separability.

\vspace{-9pt}
\section{Conclusions}\label{sec:disc}

The hybrid QGNN matches standalone DynEdge --- the strongest classical baseline --- on $F_1$ (DynEdge ahead on AP/AUC) and runs on \texttt{ibm\_fez} without error mitigation at $F_1{=}0.915$ (recalibrated). This makes HQ-RCA a hardware-executable, quantum-assisted root-cause pipeline rather than a quantum advantage. The choice is metric-driven: in this setting, the hybrid head is operationally justified when deployment optimises thresholded detection ($F_1$) --- e.g., binary escalation of an alarm cluster --- where the VQC outperforms a parameter-matched classical head (\S\ref{sec:dea}); ranked alarm triage instead favours standalone DynEdge (AP/AUC). \textit{Future work}: other embedding families; a new domain, mainframe transactions.

\vspace{-9pt}
\section*{Acknowledgments}
\enlargethispage{8\baselineskip}
The authors thank Mariagrazia Silvestro and Paolo Bandini for valuable discussions. No specific funding was received. Claude (Anthropic) assisted with the LaTeX/TikZ code for the figures.


\vspace{-10pt}
\begin{thebibliography}{00}
\setlength{\itemsep}{-1pt}
\scriptsize

\bibitem{abbasi2022}
R.~Abbasi \textit{et al.} (IceCube Coll.), ``Graph neural networks for low-energy event classification \& reconstruction in IceCube,'' \emph{JINST}, vol.~17, P11003, 2022. \href{https://arxiv.org/abs/2209.03042}{arXiv:2209.03042}.

\bibitem{cerezo2021}
M.~Cerezo \textit{et al.}, ``Variational quantum algorithms,'' \emph{Nat. Rev. Phys.}~3, 625, 2021. \href{https://arxiv.org/abs/2012.09265}{arXiv:2012.09265}.

\bibitem{elhag2024}
H.~Elhag \textit{et al.}, ``Quantum convolutional neural networks for jet images classification,'' \href{https://arxiv.org/abs/2408.08701}{arXiv:2408.08701}, 2024.

\bibitem{funcke2021}
L.~Funcke \textit{et al.}, ``Dimensional expressivity analysis of parametric quantum circuits,'' \emph{Quantum}, vol.~5, p.~422, 2021. \href{https://arxiv.org/abs/2011.03532}{arXiv:2011.03532}.

\end{thebibliography}
\end{document}